\documentclass[
reprint,
 amsmath,amssymb,
 aps,
]{revtex4-2}

\usepackage{graphicx}
\usepackage{dcolumn}
\usepackage{bm}
\usepackage{hyperref}
\usepackage[mathlines]{lineno}

\DeclareMathOperator{\Tr}{Tr}

\usepackage[usenames,dvipsnames]{xcolor}

\begin{document}

\preprint{APS/123-QED}

\title{Fluctuation-induced dynamics of nematic topological defects}

\author{Lasse Bonn$^1$}
\thanks{These authors contributed equally to this work}
\author{Aleksandra Arda\v{s}eva$^1$}
\thanks{These authors contributed equally to this work}
\author{Romain Mueller$^2$}
\author{Tyler N. Shendruk$^3$}
\author{Amin Doostmohammadi$^1$}
\thanks{Corresponding author}
\email{doostmohammadi@nbi.ku.dk}
\affiliation{$^1$Niels Bohr Institute, University of Copenhagen, Blegdamsvej 17, Copenhagen, Denmark\\$^2$Rudolf Peierls Centre for Theoretical Physics, University of Oxford, UK\\$^3$School of Physics and Astronomy, The University of Edinburgh, Peter Guthrie Tait Road, Edinburgh, EH9 3FD, UK}


\begin{abstract}
Topological defects are increasingly being identified in various biological systems, where their characteristic flow fields and stress patterns are associated with continuous active stress generation by biological entities. Here, using numerical simulations of continuum fluctuating nematohydrodynamics we show that even in the absence of any specific form of active stresses associated with self-propulsion, mesoscopic fluctuations in either orientational alignment or hydrodynamics can independently result in flow patterns around topological defects that resemble the ones observed in active systems. Our simulations further show the possibility of extensile- and contractile-like motion of fluctuation-induced positive half-integer topological defects. Remarkably, isotropic stress fields also reproduce the experimentally measured stress patterns around topological defects in epithelia. Our findings further reveal that extensile- or contractile-like flow and stress patterns around fluctuation-induced defects are governed by passive elastic stresses and flow-aligning behavior of the nematics.

\end{abstract}

\maketitle
There is growing evidence of the emergence of liquid crystalline features in biological systems (see~\cite{doostmohammadi_physics_2021} for a recent review). Significant among others is the nematic orientational order, manifest in the form of collective alignment along particular axes~\cite{marchetti_hydrodynamics_2013}, which is observed in subcellular filaments~\cite{butt_myosin_2010, sanchez_spontaneous_2012, zhang_interplay_2018, maroudas-sacks_topological_2021}, bacterial biofilms~\cite{dellarciprete_growing_2018,you2018geometry,copenhagen_topological_2021}, and cell monolayers~\cite{saw_topological_2017, duclos_topological_2017}. Due to the head-tail symmetry of the nematic particles, the lowest energy defects are of topological charge $\pm 1/2$, meaning that as one traces a loop around the defect, the particles rotate by $\pm \pi$. 
Topological defects in nematics have recently been found to be at the core of many biological functions, e.g. cell extrusion in mammalian epithelia~\cite{saw_topological_2017}, neural mound formation~\cite{kawaguchi_topological_2017}, and limb origination in the simple animal \textit{Hydra}~\cite{maroudas-sacks_topological_2021} (see~\cite{shankar_topological_2020, bowick_symmetry_2021, doostmohammadi_physics_2021} for recent reviews on physical and biological significance of topological defects).

What sets these biological nematics apart from their passive counterparts is the presence of activity: each constituent element of living matter is capable of producing work and injecting energy locally by means of active stress generation~\cite{ramaswamy_mechanics_2010}.
While the existence of quasi long-range order has also been proven for active nematics~\cite{mishra_dynamic_2010, shankar_low-noise_2018}, the dynamical properties of defects are expected to be different from the passive case since as a consequence of activity the $+1/2$ defects with polar symmetry can self-propel and move along their axis of symmetry~\cite{doostmohammadi_active_2018}.
These comet-shaped $+1/2$ defects are characterized by a head region where the director field predominantly bends and a tail region where splay is dominant~\cite{vromans_orientational_2016,tang_orientation_2017}.
For an extensile active stress, which extends along the elongation direction of active particles, the resulting motion of $+1/2$ defects is along the head, while the opposite holds for contractile active stresses~\cite{doostmohammadi_active_2018}.
This persistent movement, both in the direction of the head and the tail of the $+1/2$ defect, has been observed in various biological systems, e.g. contractile in fibroblasts~\cite{duclos_topological_2017} and extensile in epithelial monolayers~\cite{saw_topological_2017}. 
Recently, it has even been shown that perturbing the adhesion between cells can result in a switch between extensile and contractile behaviors in epithelial cell layers~\cite{balasubramaniam_investigating_2021}.
While the emergence of extensile or contractile behavior of topological defects has been widely associated to the activity of these systems, here we show that fluctuations can lead to similar patterns of flows around topological defects and result in both extensile and contractile defect behavior.

In passive nematics, fluctuations are known to drive the Berezinskii–Kosterlitz–Thouless (BKT) transition, in which spontaneously generated topological defects unbind to break the quasi long-range order~\cite{chaikin_principles_1995}.
This has been analytically shown for a $2$-dimensional passive, dry nematic, by renormalization group analyses~\cite{stein_kosterlitz-thouless_1978}.
Computational studies have shown the BKT transition for $2$D passive nematics with a lattice model with finite size scaling~\cite{vink_isotropic--nematic_2009}, and for a dry, freely moving, particle-based model for various length to width ratios~\cite{frenkel_evidence_1985, bates_phase_2000}. 
A similar BKT type transition was also reported in a discrete model of active nematics~\cite{chate_simple_2006}.

Drawing analogies with the BKT transition in passive nematics, it has been shown that in over-damped active nematics, where hydrodynamic flows are dominated and suppressed by frictional screening, self-propulsion of $+1/2$ topological defects can lead to the defect pair unbinding, destroying any (quasi) long-range orientational order~\cite{shankar_defectunbinding_2018}.
Introducing fluctuating forces coupled to the nematic alignment field, it was lately shown that such specific fluctuations can result in an effective extensile stresses in passive nematics~\cite{vafa_fluctuations_2021}. 
More recently, combining discrete, vertex-based, simulations of model cellular layers with analytical treatment of linearized nematohydrodynamics equations, it has been argued that any fluctuations can result in the appearance of ``active'' extensile or contractile nematics, depending on the flow-aligning behavior of the particles~\cite{killeen_polar_2022}.
Similarly, cell shape fluctuations in a cell-based, phase-field model of cell monolayer have been shown to affect self-propulsive features of topological defects~\cite{zhang_active_2021}.
Notwithstanding these recent works, the dynamics and flow features of topological defects in the presence of fluctuations remain poorly understood.
Moreover, it is not clear how different sources of fluctuations in hydrodynamic flows and in particle alignment affect the creation, annihilation, and motion of topological defects and whether fluctuations alone can explain experimental observations of contractile- and extensile-like defect motions in cellular layers.

Using a numerical implementation of hydrodynamic- and orientational fluctuations in a hybrid lattice Boltzmann simulation, we investigate the effects of fluctuations on a continuum nematohydrodynamics representation of nematic liquid crystals. It is important to emphasize that we do not study temperature as of the thermal fluctuations associated with the Brownian motion of the molecules, but rather mesoscopic fluctuations of the mechanical traits of the cells at the scale of the cell. Therefore, in studying fluctuations we are interested in (i) the diffusive fluctuation of the nematic director of the cells that can, for example, be caused by fluctuations in cell shape and cell alignment, and (ii) fluctuations in the forces those cells exert on their neighbors and the underlying substrate.\\ 

\noindent{\bf Model.} We employ a $2$-dimensional continuum nematohydrodynamic model~\cite{thampi_active_2016, doostmohammadi_active_2018,kos2019mesoscopic}. 
The nematic tensor order parameter, $\mathbf{Q}$, and the velocity field $\vec{v}$, evolve according to Beris-Edwards equations, and generalized incompressible Navier-Stokes equations, respectively:
\begin{eqnarray}
    \partial_t \mathbf{Q} + \vec{v} \cdot \nabla \mathbf{Q} - \mathbf{S} &=& \Gamma \mathbf{H}+\boldsymbol{\xi}^{Q}\\
    \rho(\partial_t \vec{v} + \vec{v} \cdot \nabla \vec{v}) &=& \nabla \cdot \boldsymbol{\Pi}+ \nabla\cdot\boldsymbol{\xi}^{u}, \quad \nabla \cdot \vec{v} = 0, \label{navstok}
\end{eqnarray}
where $\mathbf{H}$ is the molecular field, describing the relaxation towards minimum of the free energy that includes Landau-de Gennes bulk free energy plus the Frank elastic free energy.
The rotational diffusivity, $\Gamma$, controls the relaxation.
$\mathbf{S}$ is the co-rotation term, which captures the particle response to the gradient of flow and is a function of the flow-aligning parameter, $\xi$. 
In the momentum equation $\rho$ is the density and $\boldsymbol \Pi$ is a general stress term that includes pressure, viscous and elastic stresses, defined as $\Pi^\text{pressure}_{ij} = -p\delta_{ij}$, $\Pi^\text{viscous}_{ij} = 2\eta E_{ij}$, where $\eta$ is the dynamic viscosity and $E_{ij}$ is the rate of strain tensor, and $\Pi^\text{elastic}_{ij} = 2\xi(Q_{ij}+\delta_{ij}/2)(Q_{lk}H_{kl}) - \xi H_{ik}(Q_{kj}+\delta_{kj}/2)- \xi (Q_{ik}+\delta_{ik}/2)H_{kj} - \partial_i Q_{kl}\frac{\delta \mathcal{F}}{\delta\partial_j Q_{lk}} + Q_{ik}H_{kj} - H_{ik}Q_{kj}$. 
The effect of elastic stress on the momentum conservation, known as backflow~\cite{kos2020field}, has been numerically and experimentally shown to be relevant to $\pm 1/2$ defect annihilation dynamics in passive nematics \cite{toth_hydrodynamics_2002, blanc_dynamics_2005} (see \ref{sec:Nematohydronamicsequations} for detailed description of the governing equations).

Fluctuations in the order parameter and the momentum equations are described, respectively, as:
\begin{eqnarray}
\langle \xi^{Q}_{ij}(\vec{x},t) \xi^{Q}_{kl}(\vec{x'},t') \rangle &=& 2 k_{B}T^{Q}\Gamma \mathcal{J}_{ijkl} \delta (\vec{x}-\vec{x'})\delta (t-t'),\\
\langle \xi^{u}_{ij}(\vec{x},t) \xi^{u}_{kl}(\vec{x'},t') \rangle &=& 2 k_{B}T^{u}\eta \mathcal{J}_{ijkl} \delta (\vec{x}-\vec{x'})\delta (t-t'),
\end{eqnarray}
with zero mean. Here, the operator $\mathcal{J}_{ijkl} = \delta_{ik}\delta_{jl} + \delta_{il}\delta_{jk} -\frac{2}{d} \delta_{ij}\delta_{lk}$, with $d$ the dimension of space, renders its tensor operand symmetric and traceless~\cite{bertin_mesoscopic_2013}. 
We follow~\cite{adhikari_fluctuating_2005} for implementation of momentum conserving mesoscopic fluctuation at the lattice level and thus the fluctuations in velocity field are absorbed under a stress term (see \ref{sec:hydr_fluc} for details of the implementation).
With this formulation the amplitudes of fluctuations in both order parameter and velocity field are expressed in units of $k_{B}T$, they can be tuned independently, and more importantly setting $T^{Q}=T^{u}$ will result in mesoscale fluctuations that satisfy fluctuation-dissipation relations~\cite{hohenberg1977theory,gonnella1999phase,thampi2011lattice,bhattacharjee2008numerical}.
By setting $T^{Q} \neq T^{u}$ the fluctuations in the momentum equation are not correlated with the orientational fluctuations and in this study we first vary them independently to show that either form can result in experimentally-observed flow and stress patterns around topological defects.
To facilitate comparison with experiments, dimensionless orientational and hydrodynamic fluctuation strengths, $\hat{Q}_{k_B T}$ and $\hat{u}_{k_B T}$, are defined, respectively. To this end, inspired by the experimental characterization of the effective temperature in confined fibroblast cells~\cite{duclos_topological_2017}, we define dimensionless fluctuation strength in units of the elastic constant, $K$: $\hat{Q}_{k_B T}=k_BT^{Q}/K$ and $\hat{u}_{k_B T}=k_BT^{u}/K$.\\
\begin{figure}[tbp!]
    \centering
    \includegraphics[width=\linewidth]{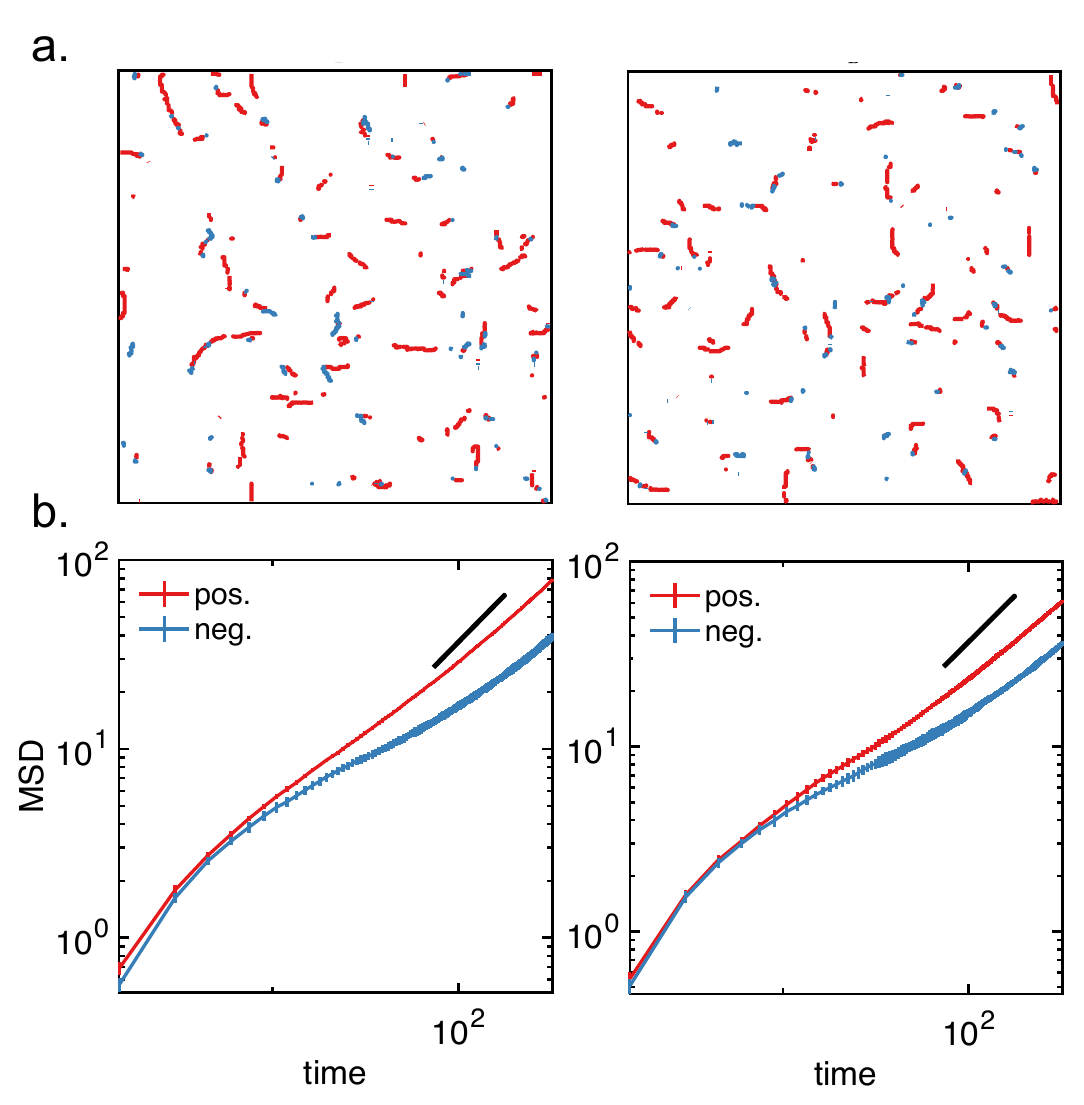}
    \caption{
    {\bf Fluctuation-induced defect kinematics.} Left and right columns compare the defect motion between the systems with hydrodynamic fluctuations ($\hat{u}_{k_B T} = 0.3$, $\xi=1$) and orientational fluctuations ($\hat{Q}_{k_B T} = 0.15$, $\xi=1$), respectively, chosen for the same topological defect density. (a) Trajectories of $+1/2$ defects (red) and $-1/2$ defects (blue) from creation time to annihilation. (b) Log-log plot of the mean-square-displacement (MSD) of defects vs time. The black line denotes slope equal to 1. Averaged over 10 realizations. The MSDs and times are nondimensionalized by the square of the characteristic coherence length scale $L_Q = \sqrt{K/A}$ and by the characteristic passive relaxation time of nematics $\tau_Q = 1/(A\Gamma)$, respectively.
    }
    \label{fig:traj}
\end{figure}

\noindent{\bf Fluctuation-induced defect kinematics.} We begin by assessing the impact of orientational and hydrodynamic fluctuations on the spatio-temporal patterns of nematic director field. Fluctuations result in the nucleation of pairs of $\pm 1/2$ topological defects. Remarkably, the nucleated pairs of defects in the presence of both hydrodynamic and orientational fluctuations show a qualitatively similar behavior to active extensile systems: after the nucleation, the $+1/2$ defect breaks away from the $-1/2$ counterpart, moving along its comet-head through the system, until it is annihilated by another $-1/2$ defect (see Supplementary Movies at~\cite{SI} for dynamics of fluctuation-induced defect motion). This persistent motion of the $+1/2$ defect is best evident in the temporal trajectory plots and the mean-square-displacement measurements of defects motion (Fig.~\ref{fig:traj}). Fluctuations lead to the emergence of distinct speeds for $+1/2$ and $-1/2$ topological defects: at short times $+1/2$ defects move faster than their $-1/2$ counterparts, while at longer times the motion of both defects becomes dominated by interactions with other defects and thus shows diffusive movement.
Such diffusive behavior for both defect types is in contrast with the propulsive $+1/2$ and diffusive $-1/2$ defect motions as observed in dense colonies of motile bacteria~\cite{meacock_bacteria_2021}.
However, the defect motion observed here for both orientational and hydrodynamic fluctuations is consistent with experimental characterization of the mean-squared-displacements of $\pm 1/2$ topological defects in the human-bronchial-cells (HBC)~\cite{blanch-mercader_turbulent_2018}, where both defect types showed diffusive behavior at long times, and suggests that the defect motions in such epithelial layers could be simply dominated by hydrodynamic fluctuations.\\
\begin{figure}[tb!]
    \centering
    \includegraphics[width=\linewidth]{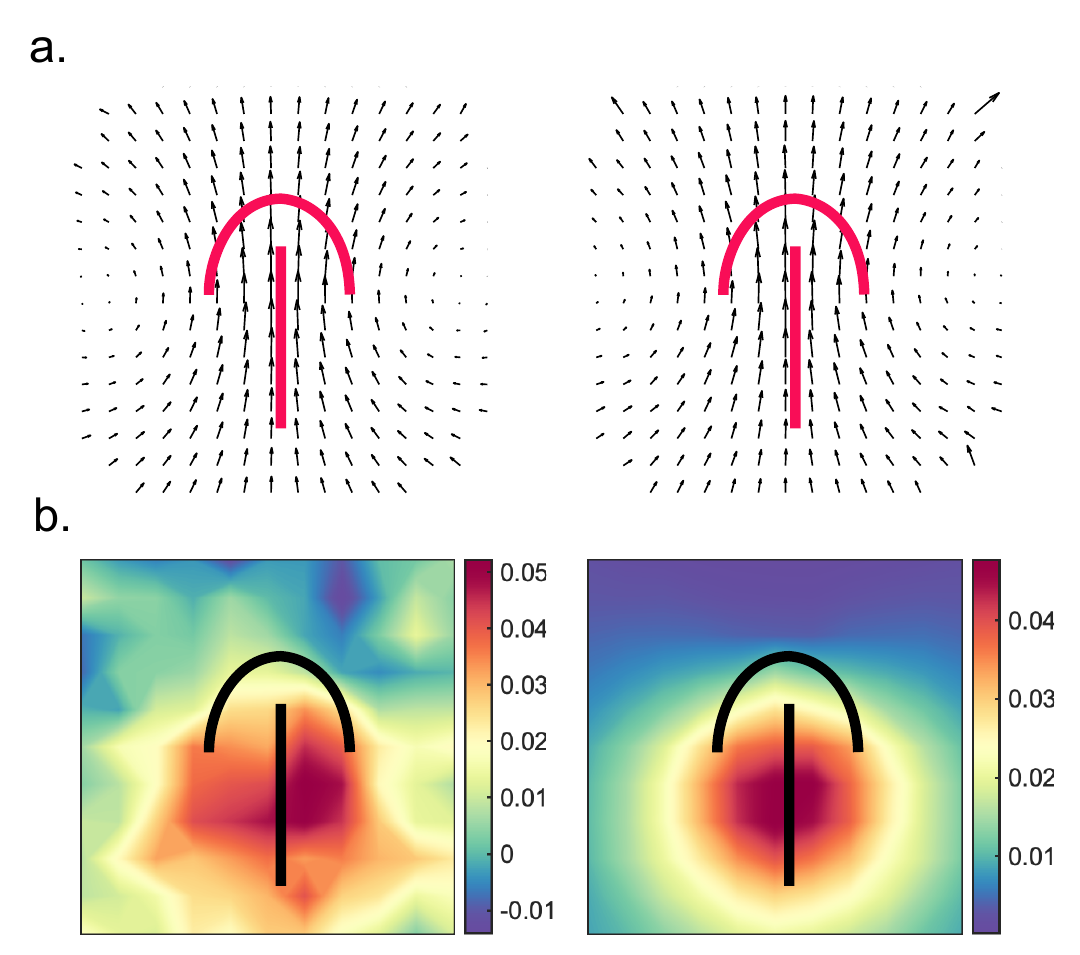}
    \caption{{\bf Fluctuation-induced defects flow and isotropic stresses.} Left and right columns compare the averaged flows and isotropic stresses between the systems with hydrodynamic fluctuations ($\hat{u}_{k_B T} = 0.3$, $\xi=1$) and orientational fluctuations ($\hat{Q}_{k_B T} = 0.15$, $\xi=1$), respectively. (a) Averaged velocity field around $+1/2$ defects. (b) Averaged isotropic stress fields around $+1/2$ defects, normalized by the bulk free energy strength $A$.
    }
    \label{fig:flow}
\end{figure}

\noindent{\bf Fluctuation-induced defects flow and isotropic stresses.} We next asked how the averaged flow fields of fluctuation-induced topological defects compare with the flow fields that have been extensively measured in experiments for different cell layers~\cite{duclos_topological_2017,saw_topological_2017,blanch-mercader_turbulent_2018} and with the theoretical predictions from active nematics~\cite{giomi_defect_2013,giomi_defect_2014}.
Interestingly, averaged flow fields around the $+1/2$ defects for both the orientational and hydrodynamic fluctuations show the typical flow jet at the defect center, pointing towards the head, accompanied by a vortex pair around the defect's axis of symmetry (Fig.~\ref{fig:flow}), which is the expected flow field for motile $+1/2$ defects in extensile active nematics~\cite{giomi_defect_2014,doostmohammadi_active_2018, meacock_bacteria_2021} and is observed in experiments on epithelial cell layers~\cite{saw_topological_2017,blanch-mercader_turbulent_2018} and neural progenitor stem cells~\cite{kawaguchi_topological_2017}.

\begin{figure}[htbp]
    \centering
    \includegraphics[width=1.0\linewidth]{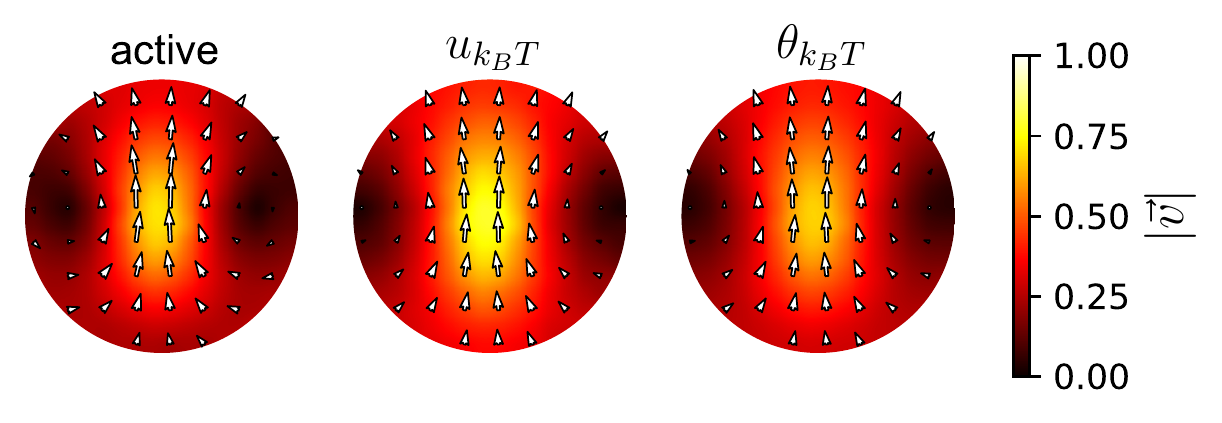}
    \caption{{\bf Fluctuation-induced versus activity-induced defect flows.} Comparison of velocity pattern and magnitude between ({\it left}) simulations with no fluctuations and active stress $\zeta=0.045$, ({\it middle}) simulations with no active stress and with hydrodynamic fluctuations $\hat u_{k_B T}=0.3$, and ({\it right}) simulations with no active stress and with orientational fluctuations $\hat \theta_{k_B T}=0.15$.
    The colorbar shows the magnitude of the velocity normalized by the characteristic passive relaxation velocity of nematics $v_Q = \Gamma \sqrt{AK}$.
    The parameters for activity and fluctuation strengths are chosen such that they result in the same defect density.
    }
    \label{fig:vel_comparison}
\end{figure}

The scales of the velocities of fluctuation-induced defects are also comparable to the ones obtained from simulations with activity. To clearly show this, we have conducted simulations with activity (and in the absence of any fluctuations) and compare the scale of the averaged velocity that is obtained around the defects to that from the simulations in the present work (no activity, and only with fluctuations).
The same velocity scales are obtained in both cases, further reinforcing the idea that fluctuation-induced features of the defects can reflect those obtained from the activity (Fig. \ref{fig:vel_comparison}).
Moreover, using estimates of the strain rates of $\sim O(10^{-2} hr^{-1})$ from experiments~\cite{saw_topological_2017}, and the correlation length of $\sim O(100 \mu m)$~\cite{balasubramaniam_investigating_2021}, and comparing them with the characteristic strain rates $\sim O(10^{-4})$ and correlation lengths $\sim O(10^{1})$ in simulation units, the velocities obtained in simulation units can be mapped to $\sim O(\mu m /hr)$ in the physical units, which are comparable to the averaged velocities around topological defects that are observed in the experiments~\cite{saw_topological_2017,balasubramaniam_investigating_2021}.\\

In addition to the characteristic flow pattern, experiments and active nematic theories have measured the isotropic stresses ($\sigma_\text{iso} = \frac{1}{2} \,\Tr(\boldsymbol{\Pi})$) around the defects to characterize the tensile and compressive forces around defect structures. 
In active nematics, an extensile defect has a region of negative isotropic stress (compression) at the head, and positive isotropic stress (tension) at the tail, inverted for a contractile defect~\cite{doostmohammadi_active_2018}.
This stress pattern has been linked to functional behavior in governing cell death and extrusion in epithelia~\cite{saw_topological_2017}. Remarkably, in addition to the flow field around $+1/2$ defects, for both orientational and hydrodynamic fluctuations, the corresponding isotropic stress patterns around the defects demonstrate the compression at the head and the tension at the tail region (Fig.~\ref{fig:flow}), demonstrating that the topological defects formed due to mesoscopic fluctuations can exhibit such experimentally observed patterns.\\

\begin{figure}[t!]
    \centering
    \includegraphics[width=\linewidth]{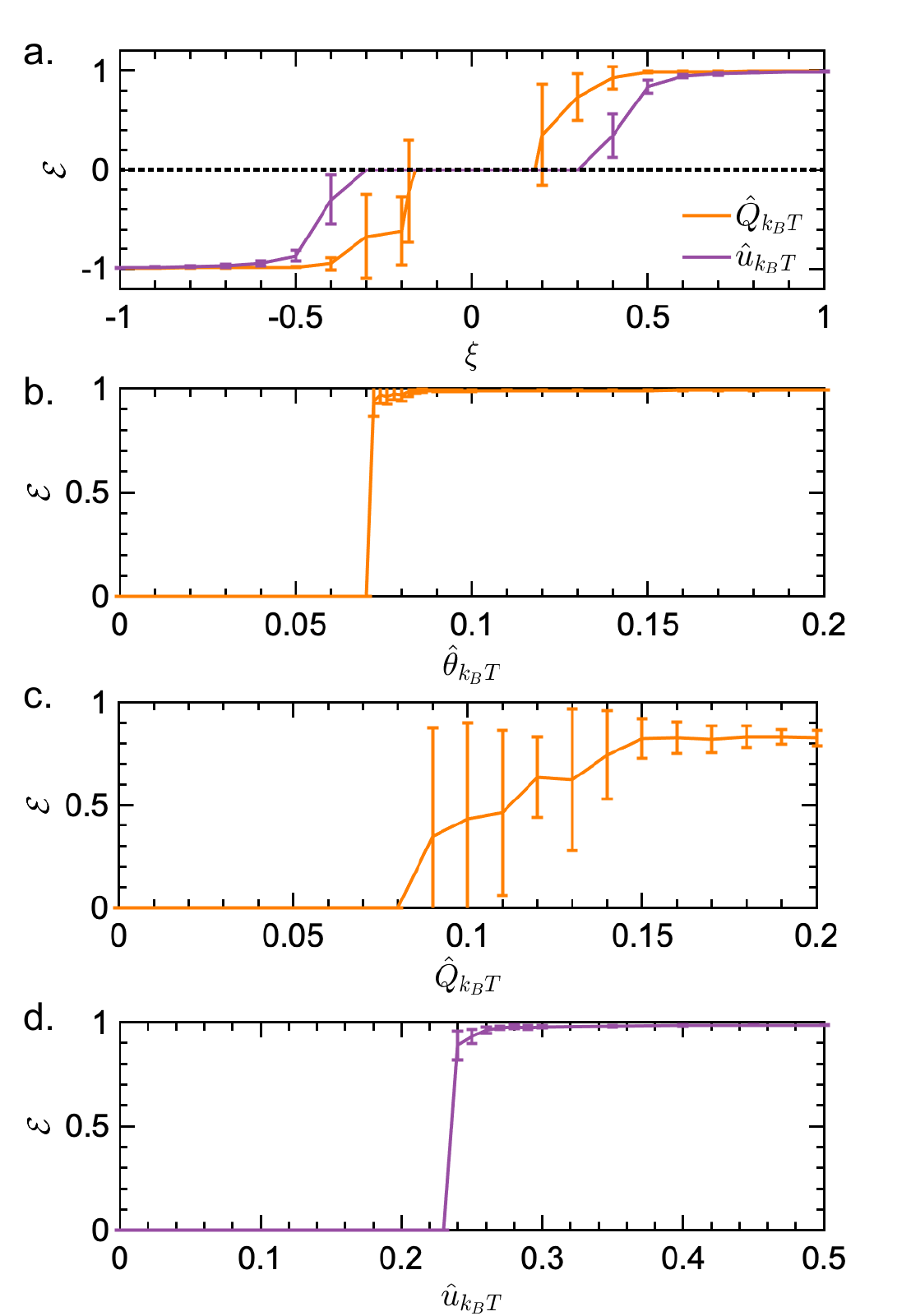}
    \caption{{\bf Fluctuation-induced extensility of $+1/2$ defects.} (a) The dependence of the extensility, $\mathcal{E}$, on the flow-aligning parameter $\xi$ for hydrodynamic fluctuations with $\hat{u}_{k_B T} = k_B T^{u} / K = 0.3$ ({\it purple line}), for orientational fluctuations with $\hat{Q}_{k_B T} = k_B T^{Q} / K = 0.15$ ({\it orange line}) and for systems without backflow ({\it black dotted line}); (b-d) The dependence of the extensility on (b) strength of the orientational fluctuation of the director angle, $\hat{\theta}_{k_B T}$, (c) nematic order parameter fluctuation strength, $\hat{Q}_{k_B T}$, and (d) hydrodynamic fluctuation strength, $\hat{u}_{k_B T}$, at $\xi = 1$.
    Averaged over $20$ realizations.
    }
    \label{fig:e1fxi}
\end{figure}

\noindent{\bf Fluctuation-induced extensility of $+1/2$ defects.} Our numerical results clearly show a tendency for the emergence of active extensile-like features around $+1/2$ defects, but it is not clear what determines such defect features.
To answer this question, we next investigated the impacts of the flow-aligning parameter and passive elastic stresses on the fluctuation-induced topological defect features, since in passive nematics the former characterizes the orientation response to flow gradients~\cite{thijssen_role_2020}, and the latter couples the orientation field to the flow~\cite{toth_hydrodynamics_2002}. 
Starting with the flow-aligning parameter, we observed disappearance of any coherent flow around defects for $\xi=0$. 
Moreover, the direction of flow around the $+1/2$ defect switches sign for negative values of the flow-aligning parameter, resembling contractile-like flows observed for monolayers of mouse fibroblasts~\cite{duclos_topological_2017} and epithelial cells with weakened cell-cell adhesion~\cite{balasubramaniam_investigating_2021}. 
To quantify the extensile- or contractile-like flow features around $+1/2$ defects we define the {\it extensility parameter}, $\mathcal{E}$, based on the averaged flow field around the defects, $\langle \vec{u}^d \rangle = (\langle u_x^d \rangle, \langle u_y^d \rangle)$, as:
 \begin{equation}\label{eq:e1}
     \mathcal{E} := \frac{\langle  u_y^\text{d} \rangle}{\langle |u^\text{d}|\rangle},
 \end{equation}
where we rotate all defects such that their comet-shaped head points in the $+y$-direction.
The extensility parameter goes from $\mathcal{E} = -1$ for a purely contractile defect over $\mathcal{E} = 0$ for no movement or isotropic movement to $\mathcal{E} = 1$ for a fully extensile defect. For both sources of fluctuations, the results show that the flow-aligning parameter, $\xi$, plays a significant role in determining the extensility of the defects (Fig.~\ref{fig:e1fxi}a).
Moreover, there is a saturation value of extensility, $\mathcal{E}$, which, for both types of fluctuations, coincides with the crossover from flow tumbling to flow-aligning behavior that is expected at $|\xi|>\frac{3q+4}{9q}$, where using approximation of the nematic order magnitude with its equilibrium value $q\sim q_{\text{eq}}=1.0$ leads to $|\xi| \gtrsim 7/9$~\cite{thijssen_binding_2020}. 
Therefore, orientation response of nematic particles to flow gradients is integral to contractile- or extensile-like behavior of $+1/2$ defects.

Not only flow-alignment is necessary for the establishment of extensile- or contractile-like flows around fluctuation-induced defects, the back coupling of the orientation to flow field through passive elastic stresses is also required.
This is evident from the results of simulations, where passive elastic stresses are turned off (Fig.~\ref{fig:e1fxi}a; {\it black solid line}). 
It is important to note that for both cases of $\xi=0.0$ and $\boldsymbol{\Pi}^\text{elastic}=0.0$ fluctuations both in orientation field and in hydrodynamics lead to defect formation, but the resulting defects do not show extensile- or contractile-like flow and stress features. 
Additionally, we numerically confirmed that the directed motion of the $+1/2$ defects is governed only by the contribution of the flow-aligning parameter, $\xi$, to the passive stress and not by the $\xi$-dependent co-rotation term in the $\mathbf{Q}$ equation. 

Next, we quantify the impact of the strength of fluctuations on the defect behavior. In addition to fluctuations in the nematic order parameter, we have further examined different implementation of the fluctuations only in the angle of the director, controlled by a dimensionless fluctuation strength $\hat{\theta}_{k_B T} = \theta_{k_B T} / K$ (see \ref{sec:alternative_orient} for details of the implementation).
As evident from Fig.~\ref{fig:e1fxi}b,c,d the extensility parameter increases from zero after the strengths of fluctuations passes above a certain threshold. This is true for all types of the fluctuations and occurs at the point where fluctuation-induced topological defects are first nucleated.\\

\begin{figure}[htbp]
    \centering
    \includegraphics[width=1.0\linewidth]{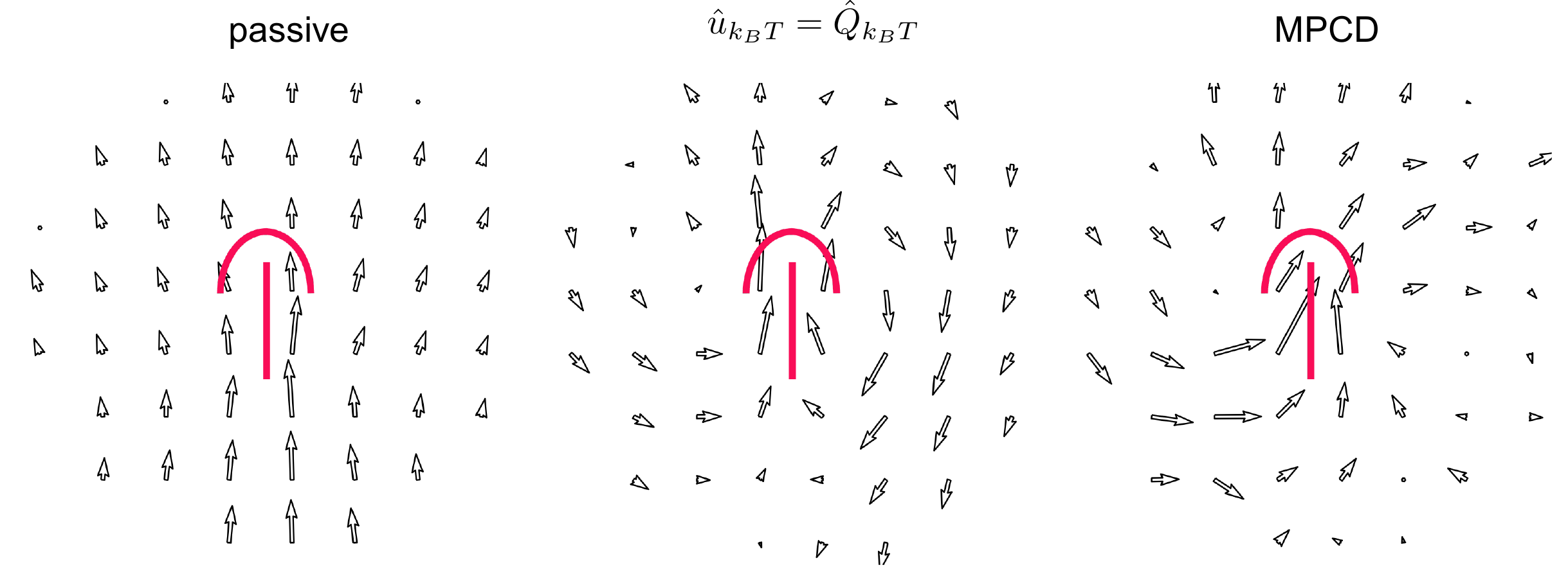}
    \caption{{\bf Fluctuation-induced defect flows in passive nematics.} Average flow field around $+1/2$ defects in passive nematics for three different scenarios:
    {\it left} results from passive nematics in the continuum model, when the system is initialized with defects and the flow around the defects is averaged during the defect life time.
    {\it middle} results from fluctuations in the continuum model satisfying fluctuation-dissipation theorem with $T^u = T^Q$.
    {\it right} results from passive nematics initialized with defects but simulated with nematic-MPCD.
    The flow pattern is robust across passive nematics, and once defects are present the characteristic extensile-like flow is set by passive elastic stresses.}
    \label{fig:pass_qu_mpcd}
\end{figure}

\noindent{\bf Fluctuation-induced defect flows in passive nematics.} Finally, to establish the governing role of the passive elastic stresses, we conduct simulations in which the noise is introduced only in the initial condition such that the initial director field contains topological defects. We then evolve this system, without any active stress and without any fluctuations, and follow the flow and stress patterns around defects as they annihilate in pairs with time until an ordered state is established. The averaged flow patterns around positive topological defects in this completely passive nematics are in agreement with both experimental observations and active nematic theories, as well as our simulation results with continuous fluctuations (Fig. \ref{fig:pass_qu_mpcd}).

Furthermore, in agreement with these results, simulations with $T^{Q}=T^{u}$ reveal similar averaged flow patterns around $+1/2$ defects (Fig. \ref{fig:pass_qu_mpcd}), even when there is no out-of-equilibrium, persistent, directed motion because the fluctuation-dissipation theorem (FDT) is satisfied.
This indicates that the extensile-like flows are set by passive elastic stresses. We further verify these $T^{Q}=T^{u}$ findings through mesoscopic simulations of passive nematohydrodynamics with inherent fluctuations via the multi-particle collision dynamics (MPCD) method~\cite{shendruk2015multi,kozhukhov2022} (see \ref{sec:mpcd} for the details of the algorithm).
MPCD results of annihilating defect pairs further corroborate the flow patterns around $+1/2$ defects (Fig. \ref{fig:pass_qu_mpcd}). Although the average flow fields around defects exhibit similar extensile-like patterns, for both MPCD and $T^{Q}=T^{u}$ simulations, the defects are short-lived and lack persistent directed motion, in contrast to fluctuations that break detailed balance (Fig.~\ref{fig:traj}a,b).
Therefore, the mechanism of extensile- or contractile-like $+1/2$ defect behavior can be understood as follows: orientational or hydrodynamic fluctuations continuously create pairs of $\pm 1/2$ topological defects; once created, passive elastic stresses generate flow fields around $+1/2$ defects that for positive (negative) values of flow-aligning parameter, $\xi$, show extensile-like (contractile-like) flow and stress features.

Our results show that the fluctuation-induced extensile- and contractile-like defects crucially depend on the sign and magnitude of the flow-aligning parameter through its contribution to the passive elastic stresses. 
The value of the flow-aligning parameter depends on the size, aspect ratio, magnitude of the order, and also interactions between the nematogens.
Indeed, the few attempts to extract the flow-aligning parameters, for the wing epithelium of \textit{Drosophila}~\cite{aigouy_cell_2010} and for mouse fibroblast cells~\cite{duclos_spontaneous_2018}, have shown that it can have a range of values and even become negative. 
Therefore it would not be surprising if different experimental systems show distinct behaviors associated with flow-tumbling or flow-aligning behavior. 
However, to our knowledge, the role of this parameter has been only marginally explored in most theoretical works and experiments.

The emergence of fluctuation-induced extensile- and contractile-like defects questions whether systems previously described as active nematics must necessarily include active stresses associated with self-propulsion.
The fluctuating forces that cells exert on their surroundings have been extensively documented \cite{plotnikov_guiding_2013, curran_myosin_2017, yang_local_2022} and are associated to various sources including, but not limited to, traction force fluctuations exerted by focal adhesion~\cite{plotnikov_force_2012, ji_fluctuations_2008, messi_traction_2020}, association/dissociation of stress fibers~\cite{guolla_force_2012}, and oscillations in Rho proteins \cite{machacek_coordination_2009, tkachenko_protein_2011}.
Importantly, these fluctuating forces are persistent and quite regular, justifying continual addition of hydrodynamic fluctuations in our model.
Similarly, fluctuations in cell shape and cell alignment have been documented~\cite{moriel_cellular_2022, zehnder_cell_2015, olenik_fluctuations_2022}.
Comparing the mesoscopic fluctuation strengths used in this study to those estimated from experiments, we find that the minimum fluctuations needed to nucleate defects are consistently smaller than experimentally estimated ones.
Force fluctuations are widely documented~\cite{plotnikov_guiding_2013}, showing the amplitude of the fluctuations to reach values as high as $50\%$~\cite{plotnikov_force_2012} in mouse embryo fibroblasts and even $10$ fold change in keratinocytes \cite{messi_traction_2020}. 
Orientational fluctuations have not been characterized as widely, although fluctuating junction lengths~\cite{curran_myosin_2017} or cell area fluctuations of up to $20\%$~\cite{zehnder_cell_2015} imply large shape changes.
However, ~\cite{duclos_topological_2017} estimated the effective temperature of $0.1 < T_\text{eff}/K<0.2$ from topological defect orientations in confined fibroblast cells, which is comparable to the values estimated from orientational fluctuations in our model.
In this regard, it is also important to note that the activity level can change across different cell types and even in the same tissue at different stages. For example, previous studies on fibroblasts~\cite{duclos_topological_2017}, have suggested that activity contribution to the defect dynamics is small and as such we conjecture that those systems could be more strongly affected by fluctuations.
Similarly, the activity levels are strongly reduced as the cells within a confluent tissue approach a glassy state, where again activity levels are significantly reduced and fluctuations can play a dominant role~\cite{malinverno_endocytic_2017,atia_geometric_2018}. Future studies should focus on effective ways for discerning the active stress-induced from mesoscopic fluctuation-induced effects in the experiments on biological matter. We finally note that our result with FDT preserving simulations in itself is very interesting and calls for extended studies of the role of mesoscale fluctuations in nematics, along the lines of the studies establishing the role of mesoscale fluctuations in capillary waves~\cite{aarts2004direct}, phase separation~\cite{gonnella1999phase}, and droplet spreading~\cite{moseler2000formation,hennequin2006drop}, that will be a focus of our future papers. It would, for example, be interesting to reexamine topological transition in passive nematic in the presence of hydrodynamics and passive elastic stresses.

\begin{acknowledgments}
A. D. acknowledges funding from the Novo Nordisk Foundation (grant No. NNF18SA0035142 and NERD grant No. NNF21OC0068687), Villum Fonden Grant no. 29476, and the European Union via the ERC-Starting Grant PhysCoMeT. A. A. acknowledges support from the European Union’s Horizon 2020 research and innovation program under the Marie Sklodowska-Curie grant agreement No. 847523 (INTERACTIONS).
\end{acknowledgments}


\setcounter{section}{0}
\renewcommand{\theequation}{\Alph{section}\arabic{equation}}
\renewcommand{\thefigure}{A\arabic{figure}}
\setcounter{figure}{0}  
\renewcommand{\thesection}{APPENDIX \Alph{section}}

\section{Nematohydronamics equations}
\label{sec:Nematohydronamicsequations}

We employ a $2$-dimensional continuum nematohydrodynamic model~\cite{thampi_active_2016, doostmohammadi_active_2018}, which we solve with the hybrid lattice Boltzmann method~\cite{marenduzzo_hlb_2007}.

The orientation of the particles is described by the director $\vec{n} = - \vec{n}$ and the order parameter is constructed $\mathbf{Q} = 2q\,(\vec{n}\vec{n} - \mathbf I/2)$ where $q$ is the strength of ordering and $\mathbf I$ is the identity tensor.
$\mathbf{Q}$ is then a symmetric, traceless tensor. The order parameter evolves according to the Beris-Edwards equation~\cite{beris_thermodynamics_1994}:
\begin{align}
    \partial_t \mathbf{Q} + \vec{v} \cdot \nabla \mathbf{Q} - \mathbf S = \Gamma \mathbf H  ,
\end{align}
where $\mathbf H$ is the molecular field and $\Gamma$ is the rotational diffusivity.
$\mathbf S$ is the co-rotation term, by which the particles respond to the gradient of flow:
\begin{eqnarray}
    \mathbf S =&&(\xi \mathbf E  + \mathbf \Omega)\cdot(\mathbf Q+\mathbf I/2) \\
    &&+ (\mathbf Q+\mathbf I/2) \cdot (\xi \mathbf E  - \mathbf \Omega) \nonumber\\
    &&-2\xi (\mathbf Q+\mathbf I/2)(\mathbf Q : \nabla \mathbf u)\nonumber,
\end{eqnarray}
where $\mathbf E = 1/2(\partial_i\vec{v}_j+\partial_j\vec{v}_i)$ is the strain rate and $\mathbf \Omega = 1/2(\partial_i\vec{v}_j-\partial_j\vec{v}_i)$ is the vorticity tensor. 
The flow alignment parameter, $\xi$, controls the alignment of the nematic director with the fluid flow, specifically tuning the relative importance of strain rate and vorticity in affecting the alignment of the director.

The molecular field, $\mathbf H$, is the negative symmetric traceless part of the derivative of the free energy, $\mathcal{F}$:
\begin{align}
    \mathbf H = -\frac{\delta \mathcal{F}}{\delta \mathbf{Q}} + \frac{\mathbf{I}}{2}\Tr\left( \frac{\delta \mathcal F}{\delta \mathbf{Q}}\right),
\end{align}
and the free energy is defined via Landau-de Gennes expansion plus an Oseen-Frank elastic term:
\begin{align}
    \mathcal F = A\left(1-\frac{1}{2}\text{Tr}(\mathbf{Q}^2)\right)^{2} + \frac{K}{2}(\nabla \mathbf{Q})^2,
\end{align}
where $A$ controls the depth of the double-well potential in the Landau-de Gennes free energy and therefore sets the energy scale of equilibrium alignment of the nematogens. 
$K$ is the Frank elastic constant used under the common one constant approximation, taking into account only divergence in $\mathbf Q$, and thus penalizing any deformations.

The velocity field, $\vec{v}$, evolves according to the incompressible Navier-Stokes equations:
\begin{align}
    \rho(\partial_t \vec{v} + \vec{v} \cdot \nabla \vec{v}) = \nabla\cdot \mathbf \Pi,\quad \nabla \cdot \vec{v} = 0 \label{navstokapp},
\end{align}
where $\rho$ is the density and $\mathbf \Pi$ is a generalized stress term.
In general, the stress can be written as a sum of pressure, viscose, elastic and active terms~\cite{doostmohammadi_active_2018}:

\begin{align}
    \Pi^\text{viscous}_{ij} &= 2\eta E_{ij}\label{eq:viscose},\\
    \Pi^\text{pressure}_{ij} &= -p\delta_{ij}\label{eq:pressure},\\
    \Pi^\text{elastic}_{ij} &= 2\xi(Q_{ij}+\delta_{ij}/2)(Q_{lk}H_{kl})\label{eq:elastic}\\
    &- \xi H_{ik}(Q_{kj}+\delta_{kj}/2)- \xi (Q_{ik}+\delta_{ik}/2)H_{kj}\nonumber\\
    &- \partial_i Q_{kl}\frac{\delta \mathcal{F}}{\delta\partial_j Q_{lk}} + Q_{ik}H_{kj} - H_{ik}Q_{kj}\nonumber,\\
    \Pi^\text{active}_{ij} &= -\zeta Q_{ij}\label{eq:active},
\end{align}
where $\eta$ is the viscosity and $p$ is the pressure.

The activity of the particles is accounted for by the active stress term (Eq. \ref{eq:active})~\cite{marchetti_hydrodynamics_2013,doostmohammadi_active_2018}.
The activity parameter, $\zeta$, can take positive or negative values which result in an extensile or contractile nematic, respectively~\cite{ramaswamy_mechanics_2010}. Unless otherwise stated the activity is set to zero in the simulations.

\begin{table}[bp]
\caption{Simulation parameters with name, symbol and value (or range) and dimension where length: $L$, mass: $M$ and time: $T$}
\label{tab:parameters}
\begin{tabular}{@{}llll@{}}
\hline
parameter                    & symbol           & value & dimension (2D)\\\hline\hline
flow-alignment          & $\xi$        & $[-1, 1]$  & $1$ \\
rotational diffusivity  & $\Gamma$         & $0.05$ & $T/M$\\
solvent viscosity  & $\eta$        & $40/6$  & $M/T$\\
density                 & $\rho$           & $40$ & $M/L^2$ \\
bulk free energy strength      & $A$              & $1$ & $M/T^2$\\
Frank elastic constant & $K$              & $0.05$  &  $ML^2/T^2$\\
numerical integration time & $\tau_{\text{LB}}$              & $1$ & $T$\\
activity & $\zeta$              & $0$ &   $M/T^2$ \\
initial noise in alignment & $n_0$     & $0.05$  & $1$ \\
velocity fluctuation    & $k_B T^u$ & $[0, 0.05]$& $ML^2/T^2$\\
director fluctuation    & $k_B T^Q$ & $[0, 0.05]$ & $ML^2/T^2$\\
director angle fluctuation    & $k_B T^{\theta}$ & $[0, 0.05]$ & $ML^2/T^2$\\
square domain length & $L_D$              & $256$   & $L$ \\\hline
\end{tabular}
\end{table}

The system is initialized in the nematic state with a director angle, $\theta_0$.
We then add an initial noise on the orientation field, $\theta$, at every lattice site in the following manner:
\begin{equation}
     \theta(t=0) = \theta_0 + n_0 \,U[-\pi/2, \pi/2],
     \label{eq:initnoise}
\end{equation} 
 with $n_0=0.05$ and $U$ the uniform distribution.

\section{Hydrodynamic fluctuations}
\label{sec:hydr_fluc}
As first noted in~\cite{adhikari_fluctuating_2005}, noise can be introduced at the level of the Boltzmann equation as 
\[
f_i(\vec r + \vec c_i, t+1) = f_i(\vec r, t) + \frac{ 1}{ \tau_{\text{LB}}} \left( f^{\text{eq}}_i(\vec r, t)-f_i(\vec r, t)\right) + \eta_i,
\]
where $\eta_i$ are correlated noises whose form must be set by the fluctuation-dissipation relation.
We consider here the D2Q9 model with a single relaxation time and have set the unit time step to unity.
Our choice for the direction vectors is
\begin{align*}
    \vec c_1 &= (0, 0), \quad \vec c_2 = (1, 0), \quad \vec c_3 = (-1, 0), \\
    \vec c_4 &= (0, 1), \quad  \vec c_5 = (0, -1), \quad \vec c_6 = (1, 1), \\
    \vec c_7 &= (-1, -1), \quad \vec c_8 =  (-1, 1), \quad \vec c_9 = (1, -1),
\end{align*}
together with the corresponding weights
\begin{align*}
w &= \left( \frac{4}{9},\frac{1}{9},\frac{1}{9},\frac{1}{9},\frac{1}{9},\frac{1}{36},\frac{1}{36},\frac{1}{36},\frac{1}{36}\right).
\end{align*}
This defines our model unambiguously and reflects exactly the conventions used in the 2D code.

As shown in~\cite{adhikari_fluctuating_2005, dunweg_statistical_2007}, the fluctuation-dissipation relation is diagonal in \emph{moment} space defined by
\begin{align*}
    m_i   = \sum_{j} \sqrt{\mu \rho w_i}\, e_{ij} f_j, \quad
    \xi_i = \sum_{j} \sqrt{\mu \rho w_i}\, e_{ij} \eta_j,
\end{align*}
where $\mu = k_B T / c_s^2$, and $c_s=1/\sqrt{3}$ is the speed of sound.
There are many possible choices for the definition of the transformation matrix, $e_{ij}$, and we follow~\cite{dunweg_statistical_2007}.
With our convention we get
\begin{align*}
\vec e_1 &= (\frac{2}{3},\frac{1}{3},\frac{1}{3},\frac{1}{3},\frac{1}{3},\frac{1}{6},\frac{1}{6},\frac{1}{6},\frac{1}{6}), \\
\vec e_2 &= (0,\frac{1}{\sqrt{3}},-\frac{1}{\sqrt{3}},0,0,\frac{1}{2 \sqrt{3}},-\frac{1}{2 \sqrt{3}},-\frac{1}{2 \sqrt{3}},\frac{1}{2 \sqrt{3}}), \\
\vec e_3 &= (0,0,0,\frac{1}{\sqrt{3}},-\frac{1}{\sqrt{3}},\frac{1}{2 \sqrt{3}},-\frac{1}{2 \sqrt{3}},\frac{1}{2 \sqrt{3}},-\frac{1}{2 \sqrt{3}}), \\
\vec e_4 &= (-\frac{2}{3},\frac{1}{6},\frac{1}{6},\frac{1}{6},\frac{1}{6},\frac{1}{3},\frac{1}{3},\frac{1}{3},\frac{1}{3}), \\
\vec e_5 &= (0,\frac{1}{2},\frac{1}{2},-\frac{1}{2},-\frac{1}{2},0,0,0,0), \\
\vec e_6 &= (0,0,0,0,0,\frac{1}{2},\frac{1}{2},-\frac{1}{2},-\frac{1}{2}), \\
\vec e_7 &= (0,-\frac{1}{\sqrt{6}},\frac{1}{\sqrt{6}},0,0,\frac{1}{\sqrt{6}},-\frac{1}{\sqrt{6}},-\frac{1}{\sqrt{6}},\frac{1}{\sqrt{6}}), \\
\vec e_8 &= (0,0,0,-\frac{1}{\sqrt{6}},\frac{1}{\sqrt{6}},\frac{1}{\sqrt{6}},-\frac{1}{\sqrt{6}},\frac{1}{\sqrt{6}},-\frac{1}{\sqrt{6}}), \\
\vec e_9 &= (\frac{1}{3},-\frac{1}{3},-\frac{1}{3},-\frac{1}{3},-\frac{1}{3},\frac{1}{3},\frac{1}{3},\frac{1}{3},\frac{1}{3}).
\end{align*}
Note that this basis is denoted $\hat {\vec e}_{i}$ in~\cite{dunweg_statistical_2007} and can be easily constructed from the vectors $\vec c_i$.
Another useful reference for the definition of the weight and basis in the D2Q9 model is \cite{chun_interpolated_2007}.
The basis is chosen to be orthogonal $\sum_{k} e_{ik} e_{jk}= \delta_{ij}$ and such that the conserved moments are given by
\begin{align*}
m_1 &= \rho = \sum_i f_i, \\
m_2 &= \rho u_x = \sum_i c_{ix} f_i, \\
m_3 &= \rho u_y = \sum_i c_{iy} f_i.
\end{align*}
In particular we have $e_{1i} = \sqrt{w_i}$, $e_{2i} = \sqrt{w_i} c_{ix} / c_s$, and $e_{3i} = \sqrt{w_i} c_{iy} / c_s$.

In moment space the noises $\xi_i$ are uncorrelated and satisfy the following fluctuation-dissipation relation
\[
    \langle\xi_i \xi_j\rangle = \delta_{ij} \frac{2 \tau_{\text{LB}} - 1}{\tau_{\text{LB}}^2},
\]
and $\xi_i=0$ for $i=1,2,3$.
The vanishing of the first three moments is simply dictated by density and momentum conservation.
Putting everything together, we finally obtain
\[
\eta_i = \sqrt{\frac{2 \tau_{\text{LB}} - 1}{\tau_{\text{LB}}^2}3 k_B T \rho w_i} \sum_{j=4}^9 e_{ij} \xi_j.
\]
One can check that conservation of particles and momentum is satisfied
\[
\sum_i \eta_i = \sum_i c_{ix} \eta_i = \sum_i c_{iy} \eta_i = 0.
\]

\section{Alternative implementation of orientational fluctuations}
\label{sec:alternative_orient}
As an alternative way of implementing orientational fluctuations we add fluctuations only in the angle of the director. To this end, at every timestep we add rotational noise to the order parameter.
This is done by calculating the angle of the director, $\theta$, from $\mathbf Q$ and then adding a scaled random $\theta_r$ at every lattice site, $l$:
\begin{align}
    \theta_{l}(t+1) = \theta_{l}(t)+ U[-\pi/2, \pi/2]\sqrt{\Gamma \theta_{k_B T}}.
    \label{eq:thetanoise}
\end{align}
We then recover the order parameter from the angle, $\theta$, and the order $S$.
This method conserves symmetry and tracelessness of the order parameter and results in an angle change at that point, without affecting the magnitude of the nematic order. Importantly, this alternative implementation also results in an extensile-like flow around $+1/2$ topological defects as shown in Fig.~\ref{fig:vel_comparison} and Fig.~\ref{fig:e1fxi}b.

\section{Mesoscopic simulations of defect dynamics} 
\label{sec:mpcd}
To verify the limit of equal effective temperatures for the velocity and director fluctuations ($T^{Q}=T^{u}$), we employ multi-particle collision dynamics (MPCD), a mesoscale coarse-grained algorithm that intrinsically simulates noisy dynamics. MPCD is a particle-based method that can simulate complex fluids, including viscoelastic fluids~\cite{Sahoo2019}, colloidal suspensions~\cite{Wani2022}, binary mixtures~\cite{tan2021}, ferrofluids~\cite{Ilg2022}, and passive~\cite{nshendruk_multi-particle_2015} and active~\cite{kozhukhov_mesoscopic_2022} nematics. Here, we provide a brief description of the nematic-MPCD algorithm and refer the reader for more detail to recent publications~\cite{nshendruk_multi-particle_2015,kozhukhov_mesoscopic_2022}. 

The nematic-MPCD algorithm discretizes the fluid into $N$ point particles  (labelled $i\in\{1,\ldots,N\}$), each with mass $m_i = m \ \forall \ i$, position, $\vec{r}_i(t)$, velocity,  $\vec{v}_i(t)$, and orientation, $\vec{u}_i(t)$. 
Their dynamics proceeds in two discrete steps: (i) ballistic streaming and (ii) multi-particle collisions.\\ 
\textbf{(i) Streaming step:} The particles move for a time $\delta t$ to a new position $\vec{r}_i\left(t+\delta t\right) = \vec{r}_i\left(t\right) + \vec{v}_i\left(t\right) \delta t. $\\
\textbf{(ii) Collision step:}  After each streaming period, the particles are sorted into cubic cells of size $a$ on a lattice that is randomly shifted to ensure Galilean invariance. Within each cell $c$, coarse-grained collision operations stochastically exchange momentum and orientation between particles, while conserving the local value. The momentum collision event is $\vec{v}_i\left(t+\delta t\right) = \left\langle\vec{v}\right\rangle_{c}\left(t\right) + \vec{\Xi}_{i,c}(t)$, where $\left\langle\vec{v}\right\rangle_{c}$ is the center of mass velocity of cell $c$ and $\vec{\Xi}_{i,c}$ is the collision operator. 
We choose the angular-momentum conserving Anderson thermostatted operator $\vec{\Xi}_{i,c} = \vec{\alpha}_i - \left\langle \vec{\alpha}_j \right\rangle_{c} + \left( \mathbf{\mathcal{I}}_{c}^{-1} \cdot \left[ \delta\vec{\mathcal{L}}_\text{vel} + \delta\vec{\mathcal{L}}_\text{ori} \right] \right)\times\vec{r}_i^\prime$~\cite{Gompper2007EPL,Gompper2007PRE}, where $\vec{\alpha}_i$ is a random velocity drawn from the Maxwell-Boltzmann distribution for thermal energy $k_\text{B}T$, $\left\langle \vec{\alpha}_j \right\rangle_{c}$ is the cell average and $\mathbf{\mathcal{I}}_{c} = m \sum_{j\in c}\left(r_j^{\prime2} \mathbf{I} - \vec{r}^\prime_{j}\vec{r}^\prime_{j}\right)$ is the moment of inertia relative to the centre of mass $\vec{r}_i^\prime=\vec{r}_i-\left\langle \vec{r} \right\rangle_{c}$. 
The first angular momentum term $\delta\vec{\mathcal{L}}_\text{vel}=\sum_{j \in c}\vec{r}_j^\prime \times \left( \vec{v}_j - \vec{\alpha}_j \right)$ corrects any spurious angular momentum introduced by the collision and the second $\delta\vec{\mathcal{L}}_\text{ori} = -\gamma \sum_{j \in c}\vec{u}_j \times \dot{\vec{u}}_j$ simulates nematic backflow for a viscous rotation coefficient, $\gamma$. 

Similarly, the orientational collision operator draws random directions from the local equilibrium distribution about the local director $\vec{n}_c$ as $\vec{u}_i\left(t+\delta t\right) = \vec{n}_c\left(t\right) + \vec{\eta}_{i,c}$, where $\vec{\eta}_{i}$ is drawn from the equilibrium Maier-Saupe distribution $\sim \exp{\left( U S_{c} \left[\vec{u}_i\cdot\vec{n}_c\right]^2 /k_\text{B}T \right)}$ for the local scalar order parameter, $S_{c}$, and a mean-field interaction constant, $U$~\cite{nshendruk_multi-particle_2015}. 
The orientation is coupled to gradients in the velocity through Jeffery's equation $\dot{\vec{u}}_i=\chi\left[ \vec{u}_i\cdot\mathbf{\Omega} + \xi \left( \vec{u}_i\cdot\mathbf{E} - \vec{u}_i\vec{u}_i\vec{u}_i:\mathbf{E} \right) \right]$ for the tumbling parameter, $\xi$, and hydrodynamic susceptibility, $\chi$. 
The Frank coefficients are a linear function of $ U/k_\text{B}T$~\cite{nshendruk_multi-particle_2015, Hijar2020PhysicaA} and nematic-MPCD has been shown to accurately simulate the coupling between fluctuating hydrodynamic modes at the mesoscopic level~\cite{Hijar2019FluctNoiseLett}.

The 2D nematic MPCD simulations are performed in square simulation boxes of size $300a$ with periodic boundary conditions. 
The density is 20 particles per cell. The streaming time step is $\delta t = 0.1 \tau$ in simulation units of $\tau=a\sqrt{m/k_\text{B}T}$. Six independent simulations are performed for $50\tau$ warmups and $250\tau$ runs. The mean-field interaction constant is $U=30k_\text{B}T$, the rotation coefficient $\gamma=0.01ma^2$, and the dimensionless tumbling parameter and hydrodynamic susceptibility are $\xi=2$ and $\chi=0.5$.
The Frank coefficients are a linear function of $ U/k_\text{B}T$~\cite{nshendruk_multi-particle_2015, Hijar2020PhysicaA} and nematic-MPCD has been shown to accurately simulate the coupling between fluctuating hydrodynamic modes at the mesoscopic level~\cite{Hijar2019FluctNoiseLett}. 
Initial particle speeds are drawn from the Maxwell-Boltzmann distributions and orientations are initialized isotropically. 
Thus, the nematic starts in a quenched disordered state, which orders through defect annihilation. Over the course of the Berezinskii–Kosterlitz–Thouless transition, the velocity field is measured in the vicinity of the defects. 

\bibliography{activematter2}

\end{document}